\documentclass[preprint,12pt]{elsarticle}

\usepackage{graphicx}
 
\usepackage{amssymb,amsmath,dsfont}
\newcounter{bla}

\journal{Computer Physics Communications}

\usepackage{enumitem}
\usepackage{url}
\usepackage[export]{adjustbox}

\begin{document}

\begin{frontmatter}

\title{Reply to Glos et al. on QSWalk.m package performance}

\author{Peter E. Falloon}
\author{Jingbo B. Wang\corref{author}}

\cortext[author] {Corresponding author.\\\textit{E-mail address:} jingbo.wang@uwa.edu.au}
\address{Department of Physics, The University of Western Australia, Perth, Australia}

\begin{abstract}
We address comments in a recent paper by Glos et al.\ [CPC 235 (2018) 414] regarding a slowdown in
performance of our QSWalk.m package for large matrix sizes. We show that the
underlying issue has been rectified in the most recent version of Mathematica. We recommend all users of QSWalk.m to upgrade to version 11.3 or above.
\end{abstract}

\end{frontmatter}

In a recent publication \cite{falloon2017} we presented a package, QSWalk.m, for the evaluation of Quantum Stochastic Walks (QSWs) in the Mathematica language. Subsequently, Glos et al.\ \cite{glos2018} have presented a similar package, QSWalk.jl, written in the Julia language. In their paper, Glos et al.\ pointed out that, for certain sets of parameters, our package suffered a severe performance degradation relative to theirs (see \cite{glos2018}, Figure 4). They correctly traced this degradation to the built-in Mathematica function MatrixExp.

We would like to report that, after independently testing the examples used in their paper, we have been able to verify the issue in Mathematica version 11.1. Happily, we also find that this issue has been resolved as of version 11.3. Figure 1 illustrates the performance of our QSWalk.m package on the same test case used in \cite{glos2018}. It can be seen that, in addition to a mild speed up for all system sizes, version 11.3 avoids the dramatic increase in computation time seen in version 11.1 for system sizes of 1181 and larger.

In light of this issue, we recommend that users of QSWalk.m who experience performance issues for large system sizes may benefit from upgrading to version 11.3.

\begin{figure}\label{fig:fig1}
\centering
\includegraphics[width=0.45\textwidth]{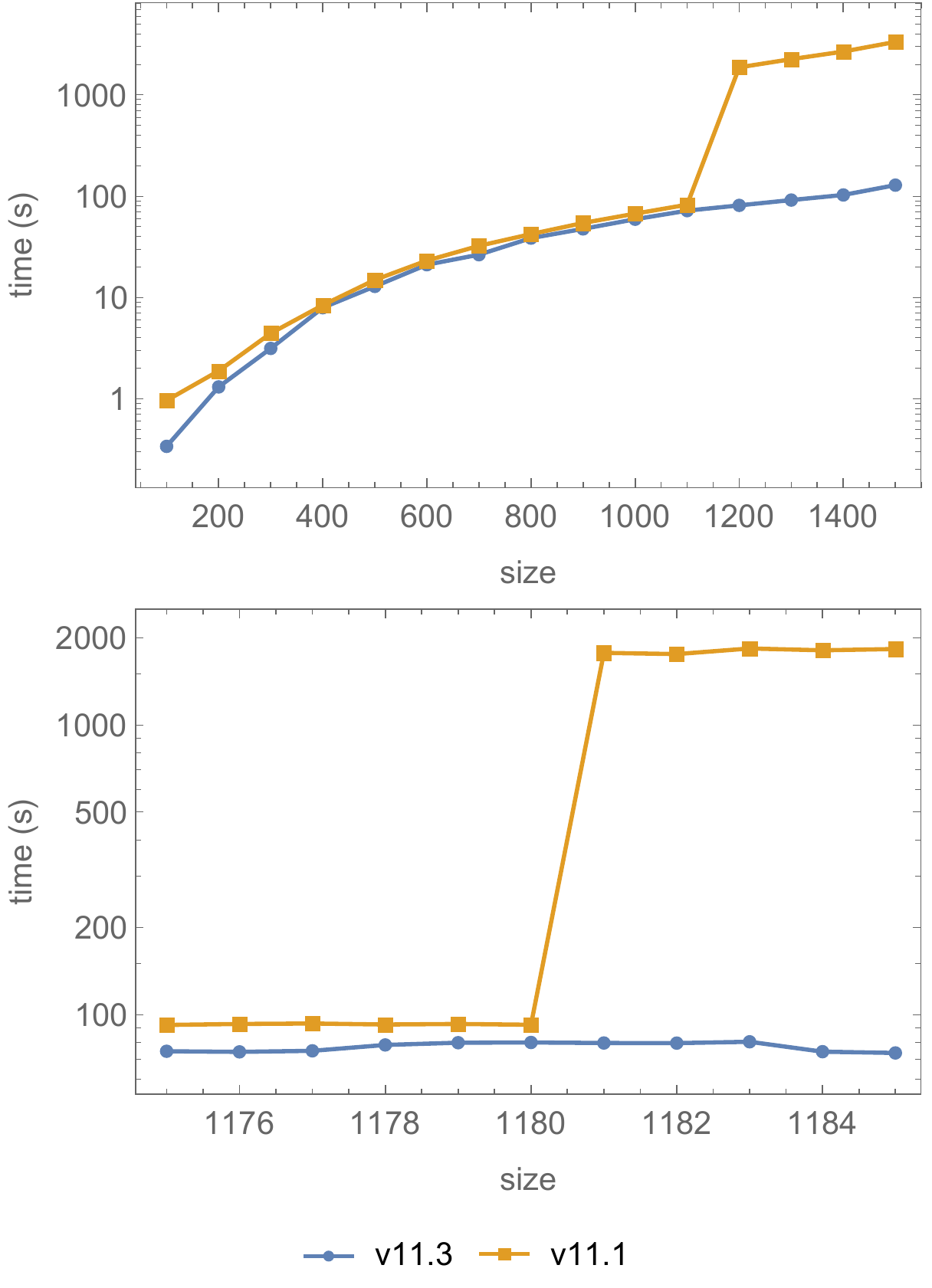}
\caption{Comparison of computation time in Mathematica version 11.1 and 11.3, for QSWs on a 1D line.}
\end{figure}

\section*{Acknowledgments}
We thank Adam Glos for kindly sharing his Mathematica code and results from \cite{glos2018}.


\section*{References}

\begin{thebibliography}{1}
\expandafter\ifx\csname url\endcsname\relax
  \def\url#1{\texttt{#1}}\fi
\expandafter\ifx\csname urlprefix\endcsname\relax\def\urlprefix{URL }\fi
\expandafter\ifx\csname href\endcsname\relax
  \def\href#1#2{#2} \def\path#1{#1}\fi

\bibitem{falloon2017}
P.~E. Falloon, J.~Rodriguez, J.~B. Wang, {QSWalk}: a {Mathematica} package for
  quantum stochastic walks on arbitrary graphs, Computer Physics Communications
  217 (2017) 162--170.

\bibitem{glos2018}
A.~Glos, J.~A. Miszczak, M.~Ostaszewski, {QSWalk.jl}: Julia package for quantum
  stochastic walks analysis, Computer Physics Communications 235 (2018)
  414--421.

\end{thebibliography}
\end{document}